\documentclass[prb,aps,twocolumn, amssymb,amsmath,floatfix,superscriptaddress]{revtex4}
\usepackage{tabularx}
\usepackage{bm}
\usepackage{euscript}
\usepackage{epsfig,psfrag,subfigure}
\usepackage{graphicx}
\usepackage{color}
\usepackage{amsfonts}
\usepackage{exscale}
\usepackage{wrapfig}
\usepackage{extarrows} 
\usepackage{placeins}
\usepackage{amsmath}
\usepackage{enumitem}

\usepackage{hyperref}
\hypersetup{
	colorlinks=true,       
    linkcolor=red,          
    citecolor=blue,        
    filecolor=magenta,      
    urlcolor=cyan           
    pdfauthor={Author},     
}

\usepackage{slashed} 
\numberwithin{equation}{section} 
\renewcommand{\theequation}{\arabic{section}.\arabic{equation}} 

\newcommand{\bea}{\begin{eqnarray}}  
\newcommand{\eea}{\end{eqnarray}}
\newcommand{\ben}{\begin{enumerate}}
\newcommand{\een}{\end{enumerate}}
\newcommand{\be}{\begin{equation}}
\newcommand{\ee}{\end{equation}}

\renewcommand{\theequation}{\arabic{section}.\arabic{equation}}

\begin{document}
\title{Fluctuating orders and quenched randomness in the cuprates} 

\author{Laimei Nie}
\affiliation{Department of Physics, Stanford University, Stanford, CA 94305, USA}

\author{Lauren E. Hayward Sierens}
\affiliation{Department of Physics and Astronomy, University of Waterloo, Ontario, N2L 3G1, Canada}
\affiliation{Perimeter Institute for Theoretical Physics, Waterloo, Ontario N2L 2Y5, Canada}

\author{Roger G. Melko}
\affiliation{Department of Physics and Astronomy, University of Waterloo, Ontario, N2L 3G1, Canada}
\affiliation{Perimeter Institute for Theoretical Physics, Waterloo, Ontario N2L 2Y5, Canada}

\author{Subir Sachdev}
\affiliation{Department of Physics, Harvard University, Cambridge, MA 02138, USA}
\affiliation{Perimeter Institute for Theoretical Physics, Waterloo, Ontario N2L 2Y5, Canada}

\author{Steven A. Kivelson}
\affiliation{Department of Physics, Stanford University, Stanford, CA 94305, USA}

\date{\today}
\begin{abstract}
We study a  quasi-2D classical Landau-Ginzburg-Wilson effective field theory in the presence of quenched disorder in which incommensurate charge-density wave and superconducting orders are intertwined. 
The disorder precludes long-range charge-density wave order, but not superconducting or nematic order. 
We select three representative sets of input parameters and compute the corresponding charge-density wave structure factors using both large-$N$ techniques and classical Monte Carlo simulations. 
Where nematicity and superconductivity coexist at low temperature, the peak height of the charge-density wave structure factor decreases monotonically as a function of increasing temperature, unlike what is seen in X-ray experiments on YBa$_2$Cu$_3$O$_{6+x}$. 
Conversely, where the thermal evolution of the charge-density wave structure factor qualitatively agrees with experiments, 
the nematic correlation length, computed to one-loop order, 
is shorter than the charge-density wave correlation length. 
\end{abstract}
\maketitle
\section{Introduction}
The cuprate superconductors manifest remarkably rich phase diagrams, but with many features common to different materials within the family \cite{kivelson03, vojta09, fradkin12}. 
In addition to the well known antiferromagnetic insulating and superconducting (SC) phases, recent experiments have 
revealed that short-range-correlated incommensurate charge-density wave (CDW) order, long known to play a prominent role in the physics of a limited subset of cuprates\cite{tranquada95, zimmermann98,hunt99, hucker11}, occurs in one way or another, in all or most cuprates\cite{hoffman02, howald03, parker10, mesaros11, wu11, ghiringhelli12, chang12, achkar12, wu13, blackburn13, neto14, comin14, achkar14, blancocanosa14, hucker14, tabis14, wu15, comin15}.  
The thermal evolution of the X-ray structure factor \cite{ghiringhelli12, chang12, achkar12, blackburn13, neto14, achkar14, blancocanosa14, hucker14} in these newly studied cases typically exhibits a gentle ``concave-upward'' onset rather than the sharp onset that is expected at the point of a thermodynamic phase transition. Meanwhile, transport\cite{ando02, daou10, cyrchoiniere15}, neutron scattering\cite{hinkov08}, STM \cite{kivelson03,lawler10, carlson11, fujita14}, and NMR\cite{wu15} measurements have revealed anisotropies that are suggestive of the existence of long-range nematicity (broken $C_4$ rotation symmetry) in an overlapping regime of the phase diagram. 

The present theoretical study is carried out with observations in the model cuprate YBa$_2$Cu$_3$O$_{6+x}$ (YBCO) in mind, specifically for a range of temperatures, $T$, low compared to the ``pseudo-gap'' crossover\cite{vojta09, fradkin12}, $T^*$,
and for a range of doping concentrations $\delta$ in the neighborhood of $\delta =1/8$ where CDW fluctuations
are experimentally detectable. In YBa$_2$Cu$_3$O$_{6.67}$ (corresponding to $\delta \approx 0.12$), the observations of Ref.~\onlinecite{chang12} then restrict
us to $T \lesssim 150$ K. We do not comment here on the higher $T$ regime where there are thermodynamic or 
spectroscopic indications of a pseudogap. We will use a fluctuating order model to study the $T$ dependence of the intertwined CDW and SC orders, and their relation to nematicity.

Previous theoretical works \cite{zachar98, sachdev04, hayward14,nie14, diamag14, wang14,maharaj14, chowdhury14} introduced classical Landau-Ginzburg models to study the competition among different order parameters in cuprates. 
In particular, Ref.~\onlinecite{hayward14} focused on the angular fluctuations of a multi-component order parameter, consisting of SC and CDW correlations in two spatial dimensions without disorder. 
Ref.~\onlinecite{nie14} investigated the effects of quenched disorder and dimensionality on CDW and nematic orders in the cuprates. 
Here, we consider a generic Landau-Ginzburg theory with a multi-component order parameter (consisting of one SC complex field $\Psi$ and two CDW complex fields $\Phi_{x,y}$) and quenched disorder in a quasi-2D system. 
We show that the $T$-dependence of the CDW structure factor depends strongly on the strength of the disorder, the dimensionality of the system, and also on other input parameters of the model. 
We also calculate other quantities such as the nematic correlation length and the integrated intensities of SC and CDW orders.

While our work was being completed, we learned of the similar analysis by Caplan {\em et al.} \cite{caplan15}.
They tune the competition between CDW and SC by an applied magnetic field, and obtain trends consistent with our results in Sec.~\ref{subsec:region1} below.

The format of this paper is as follows. 
In Sec.~\ref{sec:model}, we introduce a classical Landau-Ginzburg model for a layered system with SC and CDW orders and random-field type disorder. 
Sec.~\ref{sec:methods} illustrates the methods we use to solve this model, including the replica trick and a large-$N$ expansion, which are applied to obtain a saddle-point (mean-field) solution of the model, as well as classical Monte Carlo techniques. 
In Sec.~\ref{sec:structureFactor}, we report results for the $T$-dependence of the CDW structure factor in various regions of the phase diagram, and we discuss the effects of both disorder and dimensionality. 
We present in Sec.~\ref{sec:phaseDiagram} detailed mean-field phase diagrams as functions of various input parameters and temperature, and in 
Sec.~\ref{sec:nematic}, we show calculations of the nematic correlation length to one-loop order in the non-nematic phase.  
Finally, in Sec.~\ref{sec:discussion}, we discuss the implications of our results as well as connections to cuprate experiments.

\section{The model} \label{sec:model}
We consider a layered system with tetragonal symmetry. The charge density at position $\mathbf{r}$ in layer $m$ can be expressed as
\begin{align}
\rho_m(\mathbf{r})  &= \bar \rho + \left[ \Phi_{x,m}(\mathbf{r})e^{i\mathbf{Q_x} \cdot \mathbf{r}} +  \Phi_{y,m}(\mathbf{r})e^{i\mathbf{Q_y} \cdot \mathbf{r}}     + \text{c.c.} \right]   \nonumber \\
& \quad {} + \ldots,
\end{align}
where $\bar \rho$ is the uniform charge density, $\mathbf{Q_x}$~and~$\mathbf{Q_y}$ are incommensurate in-plane wave vectors along the $x$ and $y$ directions with equal magnitude, and $\Phi_x$~and~$\Phi_y$ are  classical CDW order parameters with slow spatial variation.
The model we study is an effective field theory of $\Phi \equiv (\Phi_x, \Phi_y)^T$ and an SC order parameter $\Psi$ in the presence of random-field disorder $h$. The corresponding classical Hamiltonian is
\bea
H &=& - \sum\limits_{\langle \mathbf{r}, \mathbf{r'} \rangle,m}  \Big[ J\Phi_m^{\dagger} (\mathbf{r}) \Phi_m(\mathbf{r'})  +K  \Psi_m^{\dagger} (\mathbf{r}) \Psi_m(\mathbf{r'}) + \text{c.c.} \Big]  \nonumber\\
&-& \sum\limits_{\mathbf{r},m} \Big[  J_z \Phi^{\dagger}_m(\mathbf{r})\Phi_{m+1}(\mathbf{r}) +V_z \Psi^{\dagger}_m(\mathbf{r})\Psi_{m+1}(\mathbf{r}) + \text{c.c.} \Big] \nonumber\\
   &-& \sum\limits_{\mathbf{r},m} J' \Big[ \Phi_m^{\dagger}(\mathbf{r}) \tau  \Phi_m(\mathbf{r}+  \mathbf{\hat x}) -  \Phi_m^{\dagger}(\mathbf{r}) \tau  \Phi_m(\mathbf{r}+ \mathbf{\hat y}) + \text{c.c.} \Big]  \nonumber\\
    &+&  \sum\limits_{\mathbf{r},m} \frac{U}{N} \Big[ |\Phi_m(\mathbf{r})|^2 + |\Psi_m(\mathbf{r})|^2   -3N\Big]^2 \nonumber\\
 &-&  \sum\limits_{\mathbf{r},m}
    \frac{\Delta}{N}     \Big[ |\Phi_{x,m}(\mathbf{r})|^2 - |\Phi_{y,m}(\mathbf{r})|^2 \Big]^2    \nonumber\\
   &+& \sum\limits_{\mathbf{r},m}  \Big[
g|\Phi_m(\mathbf{r})|^2
+ \frac{g'}{N}  |\Phi_m(\mathbf{r})|^4  \Big]   \nonumber\\
&+&\sum\limits_{\mathbf{r},m} \left[ h_m^{\dagger}(\mathbf{r})\Phi_m(\mathbf{r}) + \text{c.c.} \right] ,
 \label{completeHamiltonian}
\eea
where $|\Phi |^2 \equiv \Phi^{\dagger} \Phi$ (dagger here denotes conjugate transpose of a complex vector), 
the lattice constant is set equal to 1,  $\mathbf{r}$ and $\mathbf{r'}$ are nearest-neighbor $xy$-plane coordinates, $ \mathbf{\hat x}$ and $\mathbf{\hat y}$ are $xy$-plane unit vectors, $m$ labels the layer along $z$ direction, $N$ is the number of real components of each order parameter,
and
\be
\tau \equiv  \left( \begin{array}{cc}  
\mathbb{I}_{     N/2 \times N/2   }&   \\
 & \mathbb{-I}_{      N/2 \times N/2  }    
\end{array} \right)
\ee
 under the basis $\Big(   \Phi_x^{(1)} + i  \Phi_x^{(2)}, \ldots ,  \Phi_x^{(N-1)} + i  \Phi_x^{(N)},    \Phi_y^{(1)} + i  \Phi_y^{(2)}, \ldots ,  \Phi_y^{(N-1)} + i  \Phi_y^{(N)}     \Big) ^{T}$.
All of the following Monte Carlo results set $N=2$ (as in Refs.~\onlinecite{hayward14} and~\onlinecite{diamag14}).
The terms proportional to $J, K, J_z$ and $V_z$ are the lattice versions of the familiar gradient terms in the continuum.
Our large-$N$ and Monte Carlo calculations take $J=K$ and $J_z=V_z$. (See Sec.~\ref{subsec:disscussion_exp} for discussion of $J_z \ne V_z$.) 
Because we are interested in quasi-2D systems, we always consider the case where $1 \gg J_z/J \geq 0$.

For the special case in which $J=K$ and $J_z=V_z$, with $g=g'=J'=\Delta=0$, and in the absence of a random field, the model has a large $SO(6)$ symmetry which relates the six real components of $(\Psi,\Phi_x,\Phi_y)$.  More generally, in the absence of the random field, the model has a symmetry of $SO(2)\times SO(2) \times SO(2) \times Z_2$ , where $Z_2$ comes from invariance under $\pi/2$ spatial rotation.  The random field breaks the symmetry to a single $SO(2)$ (corresponding to the phase of the SC order parameter), although the remaining symmetries are respected on average in the disorder ensemble.

 Thermodynamic stability requires $U>0$. 
To simplify the analysis, we consider the limit $U\to+\infty$, which is  equivalent to imposing the constraint
\be
|\Phi_x|^2 +|\Phi_y|^2+  |\Psi|^2 = 3N.
\label{constraint}
\ee
\noindent  This constraint is a reflection of the experimental evidence that SC and CDW compete with each other at low temperature~\cite{ghiringhelli12, chang12, achkar12, blackburn13, achkar14, hucker14}. The Hamiltonian~\eqref{completeHamiltonian} then becomes a non-linear sigma model. 
Note, however, that calculations can be carried out in the same manner for large but finite $U$, which give qualitatively similar results.

The constraint in Eq.~\eqref{constraint} leads to an equivalency between Eq.~\eqref{completeHamiltonian} and the Hamiltonian studied by Monte Carlo methods in Refs.~\onlinecite{hayward14}~and~\onlinecite{diamag14}.  
Appendix~\ref{app:parameterConversion} provides information about how to relate parameter values used in the present and previously studied models. 
Note, however, that these previous works did not consider the effects of random-field disorder $h$ and interlayer couplings $J_z$ and $V_z$. 
Such effects will be studied in detail in this paper.
(These previous studies also excluded from the Hamiltonian the term proportional to $J'$, but the effects of this term were discussed in detail in Ref.~\onlinecite{nie14}.)

 $g$ and $g'$ determine the relative energy cost of ordering between CDW and SC. The sign of $\Delta$ distinguishes between stripe (unidirectional CDW) and checkerboard phases. In our calculation we always take $\Delta>0$ (favoring stripes). 
Note that, although it does not break $C_4$ rotational symmetry, a positive $\Delta$ favors anisotropy between CDWs along $x$ and $y$ directions.

The disorder potential $h$ is taken to be a Gaussian random field with 
\be
\overline{h_{\alpha i,m}(\mathbf{r})} = 0,
\ee
and
\be
\overline{h_{\alpha i,m}(\mathbf{r}) h_{\beta j,m'}(\mathbf{r'})} = 2\sigma^2 \delta_{\alpha \beta} \delta_{ij} \delta_{mm'} \delta(\mathbf{r}-\mathbf{r'}),
\ee
where $\overline{\cdots}$ denotes a disorder configuration average, $\alpha, \beta = x,y $ and $i,j = 1, \ldots ,N$. 
Notice that any linear couplings between $h$ and the SC order parameter $\Psi$ are forbidden by gauge invariance.  

\section{Methods} \label{sec:methods}
\subsection{Saddle-point solution in the large-$N$ limit}
We apply the replica trick\cite{dotsenko} and integrate out $h$, then decouple the quartic terms using two Hubbard-Stratonovich (HS) auxiliary fields. The resulting Hamiltonian is
\begin{align}
&H_{\text{replica}} =  \nonumber\\
&-\sum\limits_{\substack{\langle \mathbf{r}, \mathbf{r'} \rangle \\ m,a}} \Big[ 
J  \Phi_{a,m}^{\dagger} (\mathbf{r}) \Phi_{a,m}(\mathbf{r'})  
+ K \Psi_{a,m}^{\dagger} (\mathbf{r}) \Psi_{a,m}(\mathbf{r'}) + \text{c.c.} \Big] \nonumber\\
& -  \sum\limits_{\mathbf{r},m,a} \Big[  J_z\Phi^{\dagger}_{a,m}(\mathbf{r})\Phi_{a,m+1}(\mathbf{r})  + V_z\Psi^{\dagger}_{a,m}(\mathbf{r})\Psi_{a,m+1}(\mathbf{r}) + \text{c.c.}  \Big] 
 \nonumber\\
 &  -\sum\limits_{\mathbf{r},m,a} 
J' \Big[ 
\Phi_{a,m}^{\dagger}(\mathbf{r}) \tau  \Phi_{a,m}(\mathbf{r}+ \mathbf{\hat x}) 
- \Phi_{a,m}^{\dagger}(\mathbf{r}) \tau  \Phi_{a,m}(\mathbf{r}+ \mathbf{\hat y}) \Big] \nonumber\\
& {}+ \sum\limits_{\mathbf{r},m,a} i\mu_{a,m}(\mathbf{r}) \Big[   
|\Phi_{a,m} (\mathbf{r}) |^2 
+ |\Psi_{a,m}(\mathbf{r}) |^2 - 3N  \Big] \nonumber\\
& {} -\sum\limits_{\mathbf{r},m,a}  (g+ 6g')|\Psi_{a,m}(\mathbf{r})|^2 \nonumber \\
& {} + \sum\limits_{\mathbf{r},m,a} \Big[ i\eta_{a,m}(\mathbf{r}) |\Psi_{a,m}(\mathbf{r})|^2 
+ \frac{\eta^2_{a,m}(\mathbf{r}) N}{4g'} \Big] \nonumber\\
& {} +  \sum\limits_{\mathbf{r},m,a} \Bigg\{ 
  \mathcal{N}_{a,m}(\mathbf{r})  \Big[ |\Phi_{a,x,m}(\mathbf{r})|^2   -    |\Phi_{a,y,m}(\mathbf{r})|^2  \Big] 
  +  \frac{\mathcal{N}_{a,m}^2(\mathbf{r}) N}{4\Delta}  \Bigg\}   \nonumber\\
& {}  - \frac{2\sigma^2}{T} \sum\limits_{\substack{\mathbf{r},m\\  a_1,a_2}}   \Phi_{a_1,m}^{\dagger}(\mathbf{r}) \Phi_{a_2,m}(\mathbf{r})  ,
 \label{replicaHSHamiltonian}
\end{align}
where $a, a_1, a_2$ are replica indices; the HS fields $\mathcal{N}$ and $\eta$ in the second and third to last lines correspond to the quartic terms proportional to $\Delta$ and $g'$ in Eq.~\eqref{completeHamiltonian} respectively; $\mathcal{N}_m(\mathbf{r})$ is the nematic order parameter.
The Lagrange multiplier $\mu$ enforces the constraint in Eq.~\eqref{constraint}.  
The expectation values (averaged both thermally and over disorder realizations) 
of CDW and SC order parameters are obtained via diagonalization of Eq.~\eqref{replicaHSHamiltonian} in both Fourier and replica spaces. 
Assuming there is no replica symmetry breaking\cite{dotsenko}, we obtain
\bea
\overline{\langle \Phi_x^{\dagger}(\mathbf{r}) \Phi_x(\mathbf{r}) \rangle}   &=& N\int\frac{d^3\mathbf{k}}{(2\pi)^3} \left(\frac{T}{A_{\mathcal{N}}} + \frac{2\sigma^2}{A^2_{\mathcal{N}}} \right),\\
\overline{\langle \Phi_y^{\dagger}(\mathbf{r}) \Phi_y(\mathbf{r}) \rangle}   &=& N\int\frac{d^3\mathbf{k}}{(2\pi)^3} \left(\frac{T}{B_{\mathcal{N}}} + \frac{2\sigma^2}{B^2_{\mathcal{N}}} \right),\\
\overline{\langle \Psi^{\dagger}(\mathbf{r}) \Psi(\mathbf{r}) \rangle}   &=& N\int\frac{d^3\mathbf{k}}{(2\pi)^3} \frac{T}{C} ,
\eea
where
$\left\langle \ldots \right\rangle$ and $\overline{\cdots}$ denote thermal and disorder averages respectively, and
\bea
A_{\mathcal{N}},B_{\mathcal{N}} &\equiv& -2J(\cos k_x + \cos k_y) 
\mp 2J'(\cos k_x - \cos k_y) \nonumber\\
& & {} -2J_z \cos k_z \pm \mathcal{N} + \mu,\\
C &\equiv&  -2K(\cos k_x + \cos k_y) -2V_z \cos k_z \nonumber\\
& & {} -g- 6g' +\eta + \mu,
\eea
where we have taken the mean-field approximation that $\mu_m(\mathbf{r}) \equiv\mu, \ \ \mathcal{N}_m(\mathbf{r}) \equiv \mathcal{N}, \ \ \eta_m(\mathbf{r}) \equiv \eta$ for any $(m,\mathbf{r})$.
The saddle-point equations that are to be solved self-consistently are
\be
(3-\tilde m^2)N =\overline{ \langle \Phi_x^{\dagger}(\mathbf{r}) \Phi_x(\mathbf{r}) \rangle}  + \overline{\langle \Phi_y^{\dagger}(\mathbf{r}) \Phi_y(\mathbf{r}) \rangle} + \overline{\langle \Psi^{\dagger}(\mathbf{r}) \Psi(\mathbf{r}) \rangle},
\label{saddle1}
\ee
\be
-\frac{\mathcal{N}N}{2\Delta} = \overline{\langle \Phi_x^{\dagger}(\mathbf{r}) \Phi_x(\mathbf{r}) \rangle} -  \overline{\langle \Phi_y^{\dagger}(\mathbf{r}) \Phi_y(\mathbf{r}) \rangle},
\label{saddle2}
\ee
\be
\left(\frac{\eta }{2g'}-\tilde m^2\right)N =  \overline{\langle \Psi^{\dagger}(\mathbf{r}) \Psi(\mathbf{r}) \rangle} ,
\label{saddle3}
\ee
where $\tilde m $ is magnitude of the SC condensate in SC phase, and we have redefined $\eta$ and $\mu$ to absorb a factor of $i$. $\mathcal{N}$ in Eq.~\eqref{saddle2} is the mean-field nematic order parameter defined as the anisotropy between CDWs along $x$ and $y$ directions. In other words, in our theory the nematic order is a vestigial order of CDW~\cite{nie14}, breaking $C_4$ rotational symmetry but preserving lattice translational symmetry (see Sec.~\ref{subsec:disscussion_exp} for discussion of other possible origins of nematic order). We numerically solve the above equations by computing the integrals with the given lattice regularization.
\subsection{Monte Carlo}

Classical Monte Carlo methods are capable of measuring standard equilibrium thermodynamic estimators of our model, such as energy, magnetization, CDW and SC structure factors, and various correlation lengths.
Our simulations are performed on finite-size lattices and involve a combination of local\cite{metropolis, hastings, newman} and non-local\cite{wolff, newman} importance sampling techniques, as described in detail in Ref.~\onlinecite{diamag14}.
Non-local sampling is especially important at low temperatures, where both efficiency and ergodicity issues can become significant. 
Note that our non-local sampling involves a modified Wolff cluster update that is only possible when $J=K$, $J'=0$ and $J_z = V_z$ in our model. 
In all of the following plots, the large-$N$ and corresponding Monte Carlo results adopt the same input parameters and can be directly compared with each other, except that
the large-$N$ mean-field calculations set $J'=0.01J$, while the Monte Carlo calculations set $J'=0$ in order to enable cluster sampling and thus avoid non-ergodic behaviour. 
Careful studies reveal that this slight difference in parameters does not have a significant effect on the structure factors shown in our plots.

In the presence of random-field disorder ($\sigma \neq 0$), Monte Carlo calculations of the CDW structure factor $S_{\Phi_x}(\mathbf{k}=0)$ (which will be defined in Sec.~\ref{sec:structureFactor}) require averaging over many independent realizations of disorder, $\{ h_{\alpha i} \}$. 
Our numerical studies reveal that, as $\sigma$ is increased, the distribution of $S_{\Phi_x}$ over various Realizations of Disorder (ROD) becomes increasingly asymmetric due to the fact that $S_{\Phi_x}$ is a complicated, non-linear function of the disorder fields $h_{\alpha i}$. 
As a result, the average value, $\left[ \left\langle S_{\Phi_x} \right\rangle \right]_\text{ROD}$, of this distribution becomes different from its typical value, $\exp \left[ \ln \left\langle S_{\Phi_x} \right\rangle \right]_\text{ROD}$, where $\left\langle \ldots \right\rangle$ and $\left[ \ldots \right]_\text{ROD}$ denote thermal and disorder averages, respectively. However, in order to allow comparison with large-$N$ results
(for which calculations of the typical value are extremely difficult), 
all of the following Monte Carlo results correspond to  average values of the disorder distributions. The qualitative behaviour of the structure factors is similar if one instead examines the typical values.

Note that, in cases where no disorder is present ($\sigma=0$), the error bars in our Monte Carlo results correspond to thermal averaging. In the presence of disorder, error bars instead correspond to the standard deviation of the mean over independent ROD. Our results average over between $10^2$ and $10^3$ ROD.  We find that both when we increase $\sigma$ and when we study temperatures near the structure factor peak, more ROD are required in order to obtain high-quality numerical results.

Unless otherwise stated, Monte Carlo simulations are performed on lattices of size $32 \times 32 \times 8$.
By examining the behaviour of the structure factor $S_{\Phi_x}(\mathbf{k}=0)$ on larger lattices for a selective set of input parameters, we have determined that this size is generally sufficient to ensure that the data has converged within a few percent (at worst) of the infinite-size limit.


\section{CDW Structure factor}\label{sec:structureFactor}
\begin{figure}[t]
	\begin{center}
	\includegraphics[width=3in]{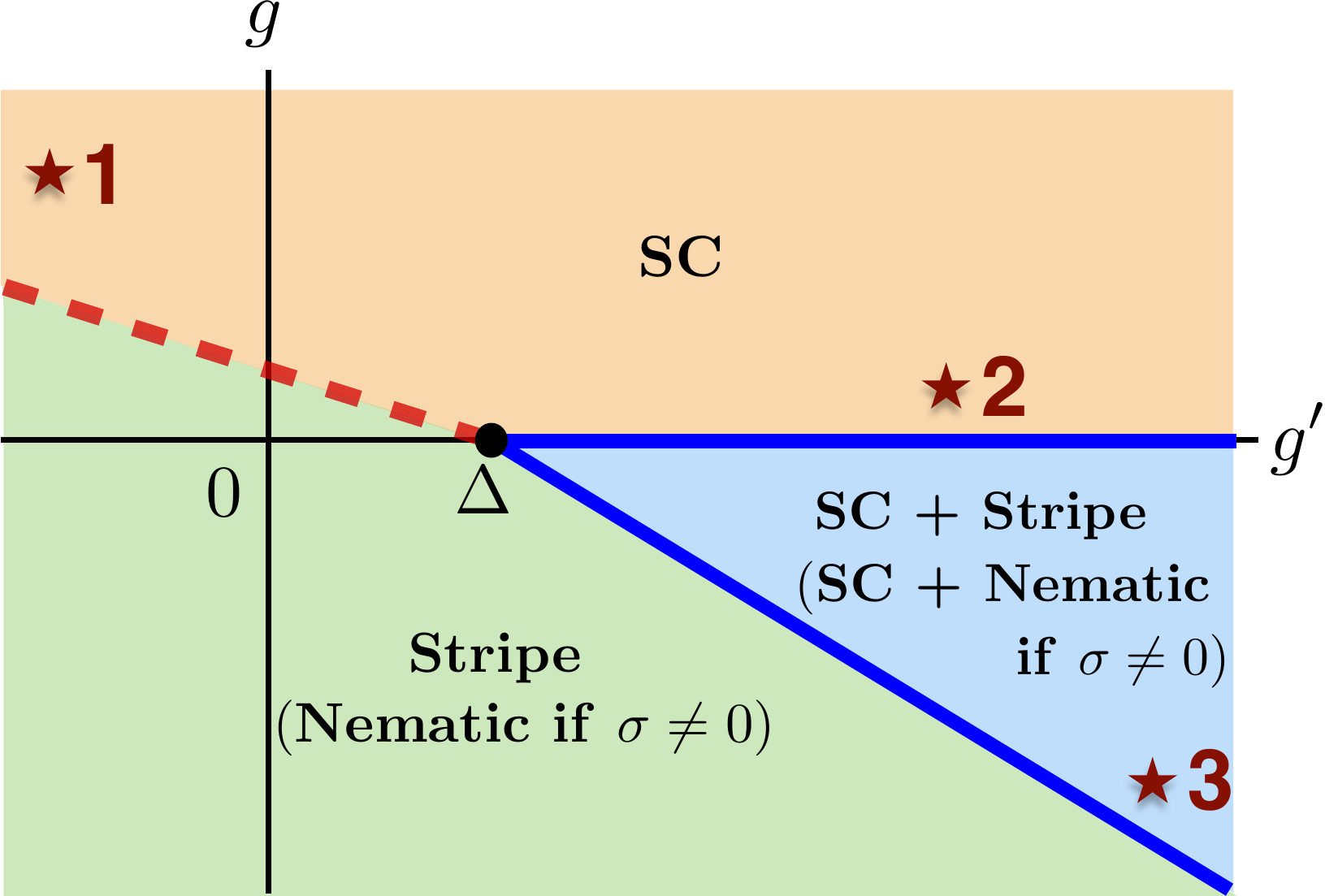}
	\end{center}
	\caption{Zero-temperature, zero-disorder phase diagram (see Eq.~\eqref{completeHamiltonian} for definitions of $g$ and $g'$).
	 For $g>0$ and $g'>0$ it is energetically expensive to form CDW order, therefore the ground state is SC; for $g<0$ and $g'<0$ the system prefers unidirectional CDW.
		The dashed and solid lines mark first- and second-order transitions, respectively. 
	A bi-critical point is located at $g'=\Delta$, $g=0$. 
	The stripe (SC+stripe) phase becomes a nematic (SC+nematic) phase in the presence of disorder. 
	The stars mark the three sets of input parameters used in our calculations. 
}
	\label{fig:phasediagram+structurefactor}
\end{figure}
Our starting point is the $T=0$ zero-disorder phase diagram shown in Fig.~\ref{fig:phasediagram+structurefactor}. 
Three phases emerge from a bi-critical point: an SC, a stripe, and a coexisting SC and stripe phase. 
For finite $T$, finite disorder and non-zero interlayer coupling, the bi-critical point and phase boundaries shift; see Sec.~\ref{sec:phaseDiagram} for a detailed discussion of the evolution of the phase diagram. 
Using both large-$N$ saddle-point methods and Monte Carlo simulations, we compute the CDW structure factor
\be
S_{\Phi_x}(\mathbf{k}=0) = \left. \frac{1}{N} \overline{\langle  \Phi_x^{\dagger}(\mathbf{k}) \Phi_x(\mathbf{k}) \rangle}  \right|_{\mathbf{k}=0}
\ee
as a function of $T$ using three sets of input parameters, as indicated by the stars in Fig.~\ref{fig:phasediagram+structurefactor}. $S_{\Phi_x}(\mathbf{k}=0)$  and  $S_{\Phi_y}(\mathbf{k}=0)$ represent X-ray scattering intensities due to CDW at wave vectors $\mathbf{Q_x}$ and $\mathbf{Q_y}$ respectively. 
In all of the following plots of $S_{\Phi_x}(\mathbf{k}=0)$ vs. $T$, Monte Carlo and large-$N$ saddle-point results agree qualitatively, but show significant quantitative differences, especially for temperatures close to those at which the CDW peak height is maximal. 
  These differences can be reduced if $1/N$ corrections are included (see Appendix~\ref{app:oneoverN} for details). 

\subsection{Region 1}\label{subsec:region1}
\begin{figure}[t]
	\begin{center}
	\includegraphics{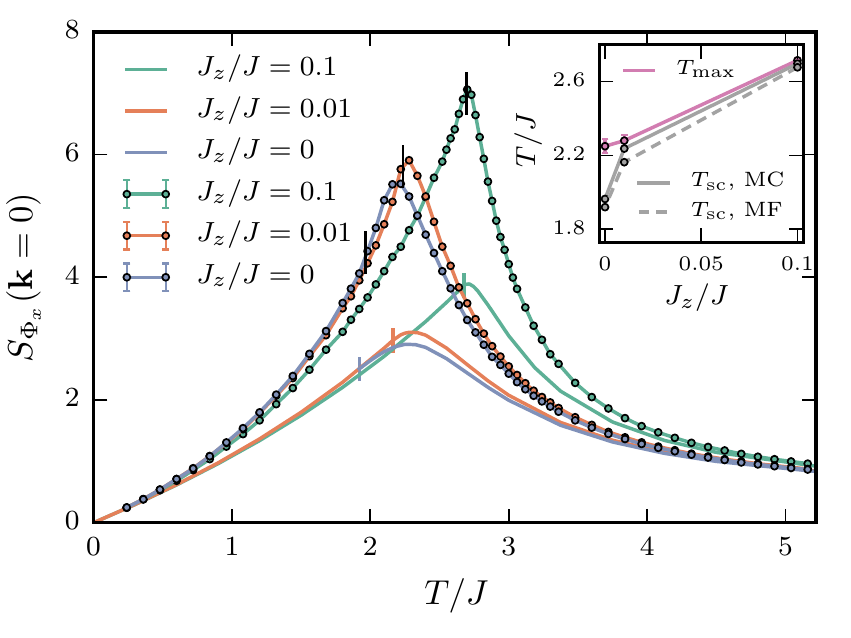}
	\end{center}
	\caption{%
	The CDW structure factor $S_{\Phi_x}(\mathbf{k}=0)$ as a function of $T$ with $\sigma=0$, increasing $J_z$, and the parameters given in Eq.~\eqref{input} in Region 1. Solid lines without indicated data points are the large-$N$ saddle-point results, while the points with error bars (which are smaller than the data points in this case) are data from Monte Carlo. 
	The lines connecting Monte Carlo data points are merely guides to the eyes.   
	Short vertical lines mark the locations of SC transition temperatures, $T_\text{sc}$. 
	For $J_z=0$, one can still calculate the SC transition temperature from the Monte Carlo data, but there is no transition at mean-field level. In this case, the short vertical line corresponding to the large-$N$ data instead marks the temperature at which the SC mass term becomes exponentially small. 
	In the inset, we demonstrate that $T_\text{sc}$ approaches $T_\text{max}$ from below as $J_z$ is increased. 
	Monte Carlo (MC) results for $T_\text{sc}$ are consistently higher than the corresponding large-$N$ mean-field (MF) results, while both MC and MF give the same estimates for $T_\text{max}$ within error.
	}
	\label{fig:region1-1}
\end{figure}

\noindent \textbf{1.  
Zero disorder ($\sigma=0$) and various  interlayer couplings, $J_z$.} The input parameters of Eq.~\eqref{completeHamiltonian} are taken to be
\begin{align}
& K=J, \:\,  J'=0.01J \text{ ($J'=0$ in Monte Carlo)},   \nonumber\\ %
&  V_z = J_z, \:\, g = 0.35J, \:\, g' = -0.033J, \:\, \Delta=0.033J. 
 \label{input}
\end{align}
This reproduces (approximately) the fitting parameters used in Ref.~\onlinecite{hayward14}:
\be
\lambda=1, \ \ w=  -0.2, \ \ \tilde g =  0.35, \ \ \tilde g' = 0.
\ee 
(See Appendix~\ref{app:parameterConversion} for definitions of $\lambda, w, \tilde g$ and $\tilde g'$.)

As shown in Fig.~\ref{fig:region1-1}, 
the CDW structure factor grows with decreasing temperature down to a non-zero $T_\text{max}$, at which point 
it attains maximum value. Below $T_\text{max}$, $S_{\Phi_x}(\mathbf{k}=0)$ decreases until it reaches zero at $T=0$. 
$S_{\Phi_x}(\mathbf{k}=0)$ is convex on both sides of $T_\text{max}$.
As the system becomes more 3D-like, the prominence of the maximum is enhanced, and the SC transition temperature $T_\text{sc}$ approaches $T_\text{max}$ (see inset of Fig.~\ref{fig:region1-1}). Large-$N$ calculations find that $T_\text{sc}$ exceeds $T_\text{max}$ when $J_z > 5J$ (not shown). 
Monte Carlo simulations for $J_z = 0$, $0.01$ and $0.1$ are performed on lattices of sizes $64 \times 64$, $32 \times 32 \times 8$, and $24 \times 24 \times 14$, respectively.
\\

\noindent \textbf{2. Disorder is present ($\sigma>0$) with fixed interlayer coupling.} We fix $J_z=0.01J$, increase the value of $g$ to $g = 0.7J$ in order to ensure that the ground state is SC, and otherwise keep  the  input parameters the same as in Eq.~\eqref{input}.
 \begin{figure}[t]
	\begin{center}
	\includegraphics{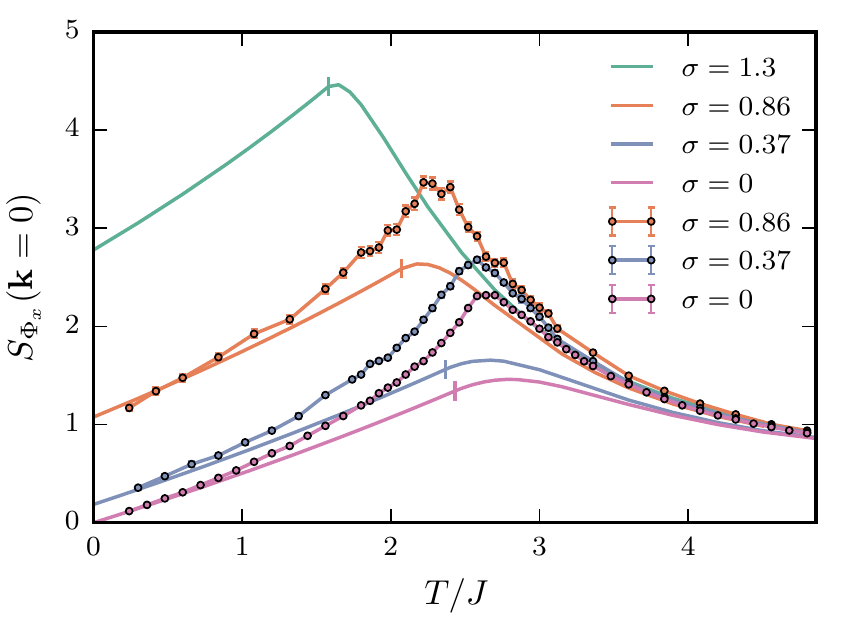}
	\end{center}
	\caption{$S_{\Phi_x}(\mathbf{k}=0)$ as a function of $T$ in Region 1 , with fixed $J_z \ne 0$ and increasing disorder, with parameters given in Eq.~\eqref{input} but with $g = 0.7$. We show both large-$N$ (undecorated lines) and Monte Carlo (points with error bars) results.
	Short vertical lines indicate SC transition temperatures.}  
\label{fig:region1-2}
\end{figure}
As shown in Fig.~\ref{fig:region1-2}, the CDW structure factor as a function of temperature still has 
a similar shape, but for $\sigma>0$ it develops a non-zero value at $T=0$, which can be understood as a consequence of the disorder-pinning effect of CDW fluctuations. 
As $\sigma$ increases, $T_\text{sc}$ and $T_\text{max}$ get closer to each other, but $T_\text{sc}$ remains smaller than $T_\text{max}$. 
\subsection{Region 2} 
\begin{figure}[t]
	\begin{center}
	\includegraphics{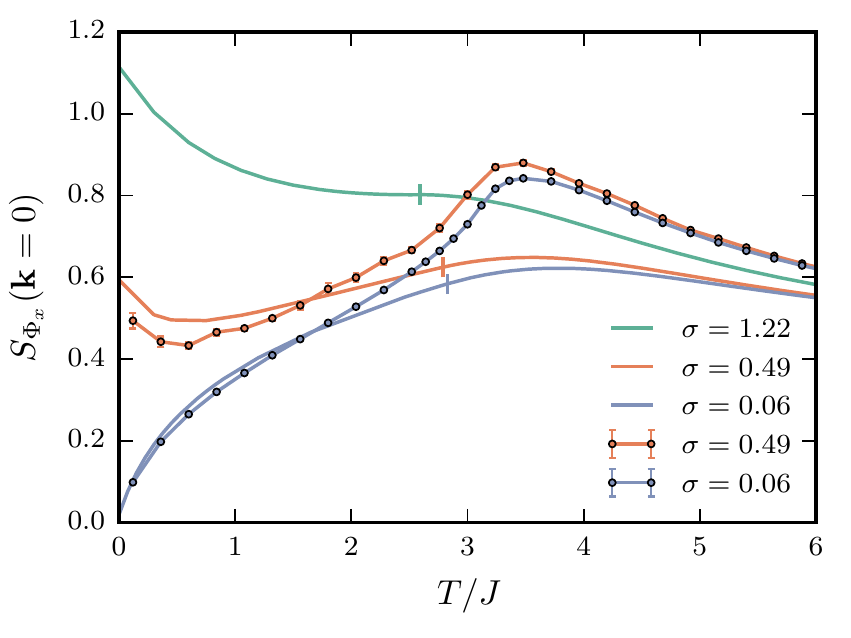}
	\caption{$S_{\Phi_x}(\mathbf{k}=0)$ as a function of $T$ in Region 2, using large-$N$ (undecorated lines) and Monte Carlo (points with error bars) methods.
	Here we fix $J_z = 0.01J$ and increase the disorder strength. Input parameters are given in Eq.~\eqref{input2}. Short vertical lines correspond to SC transition temperatures. }
	\label{fig:region2}
	\end{center}
\end{figure}

\noindent We change the values of $g'$ and $\Delta$ so that the input parameters for Eq.~\eqref{completeHamiltonian} become
\begin{align}
& K=J, \:\,  J'=0.01J \text{ ($J'=0$ in Monte Carlo)},  \nonumber\\
&  V_z = J_z=0.01J, \:\, g = 0.35J, \:\, g' = 0.533J, \:\, \Delta=0.167J.
\label{input2}
\end{align}
Although Region 2 and Region 1 both have SC as the zero-disorder ground state, their CDW structure factors behave very differently under the effect of disorder. As shown in Fig.~\ref{fig:region2}, the feature of maximum intensity at $T_\text{max}$ is suppressed by disorder, in contrast with Fig.~\ref{fig:region1-2}. Moreover, the structure factor begins to increase again as $T$ approaches zero, unlike the situation in Region 1 where the structure factor decreases monotonically as $T$ is decreased below $T_\text{max}$. In both Regions 1 and 2 the intensity at $T=0$ is enhanced by disorder, which again is a disorder-pinning effect. 
\subsection{Region 3}\label{subsec:region3}
\noindent The input parameters of Eq.~\eqref{completeHamiltonian} are taken
to be
\begin{align}
& K=J, \:\, J'=0.01J \text{ ($J'=0$ in Monte Carlo)},  \nonumber\\
&  J_z = V_z=0.01J, \:\, g = -2.8J,  \:\, g' = 0.667J, \:\, \Delta=0.167J.
\label{input3}
\end{align}
 In the absence of quenched randomness, there is a finite-temperature transition in this region to a CDW (stripe) phase, and SC and stripe order coexist at low $T$ .
In the presence of quenched randomness, no long-range CDW order occurs, but for weak enough randomness, there remain finite-$T$ transitions below which nematic order and SC develop sequentially.  
\begin{figure}[t]
	\begin{center}
	\includegraphics{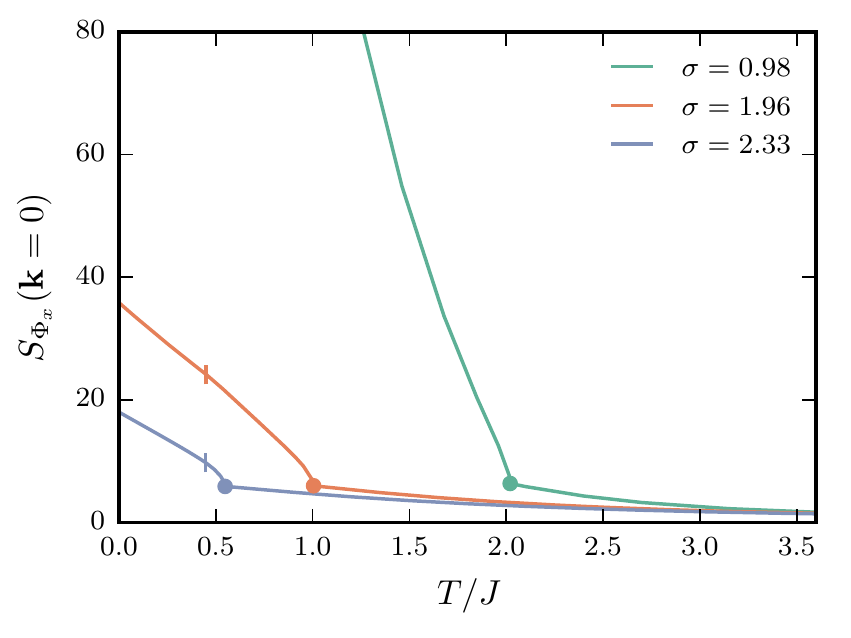}
	\end{center}
	\caption{$S_{\Phi_x}(\mathbf{k}=0)$ as a function of $T$ in Region 3, with fixed $J_z =0.01J$, increasing disorder, and parameters given in Eq.~\eqref{input3}. 
Short vertical lines and dots mark SC and nematic transition temperatures, respectively. 
$T_\text{sc}$ remains almost constant when disorder strength $\sigma$ is varied, while the nematic transition temperature depends strongly on $\sigma$.
We show only large-$N$ results here, since Monte Carlo results are difficult to obtain with these input parameters in the presence of such strong disorder. }
	\label{fig:region3}
\end{figure}

As shown in Fig.~\ref{fig:region3}, CDW correlations are greatly enhanced below the nematic transition in the
preferred direction, and these correlations grow monotonically towards $T=0$.  
In contrast, the SC transition, which occurs at a lower temperature, has very little influence on the behaviour of the CDW structure factor. 
While in Figs.~\ref{fig:region1-1},~\ref{fig:region1-2}~and~\ref{fig:region2} we found that the SC transition had a dramatic effect on CDW correlations, in this regime we find instead that nematicity plays an overwhelming role in determining the $T$-dependence of CDW correlations. 
In other words, SC and nematic transitions tend to decrease and increase CDW correlations respectively, and nematicity always wins when these two factors compete. 
We should emphasize that this is not a fine-tuning effect; as long as there is a nematic phase with a critical temperature larger than the SC transition temperature, the lack of a maximum in the thermal evolution of CDW structure factor is generally observed for a variety of input parameters. 

We did not study the region where the ground state is purely stripe in the absence of disorder (the green region in Fig.~\ref{fig:phasediagram+structurefactor}). Due to the lack of SC, this region is probably less relevant to experiments.

\section{Phase diagrams}\label{sec:phaseDiagram}
In this section we discuss in detail how the phase diagram evolves with increasing temperature and disorder.  All phase diagrams are determined by the large-$N$ saddle-point method. \\
 
\subsection{Zero temperature} 
\begin{figure}[t]
	\begin{center}
	\includegraphics[width=2.5in]{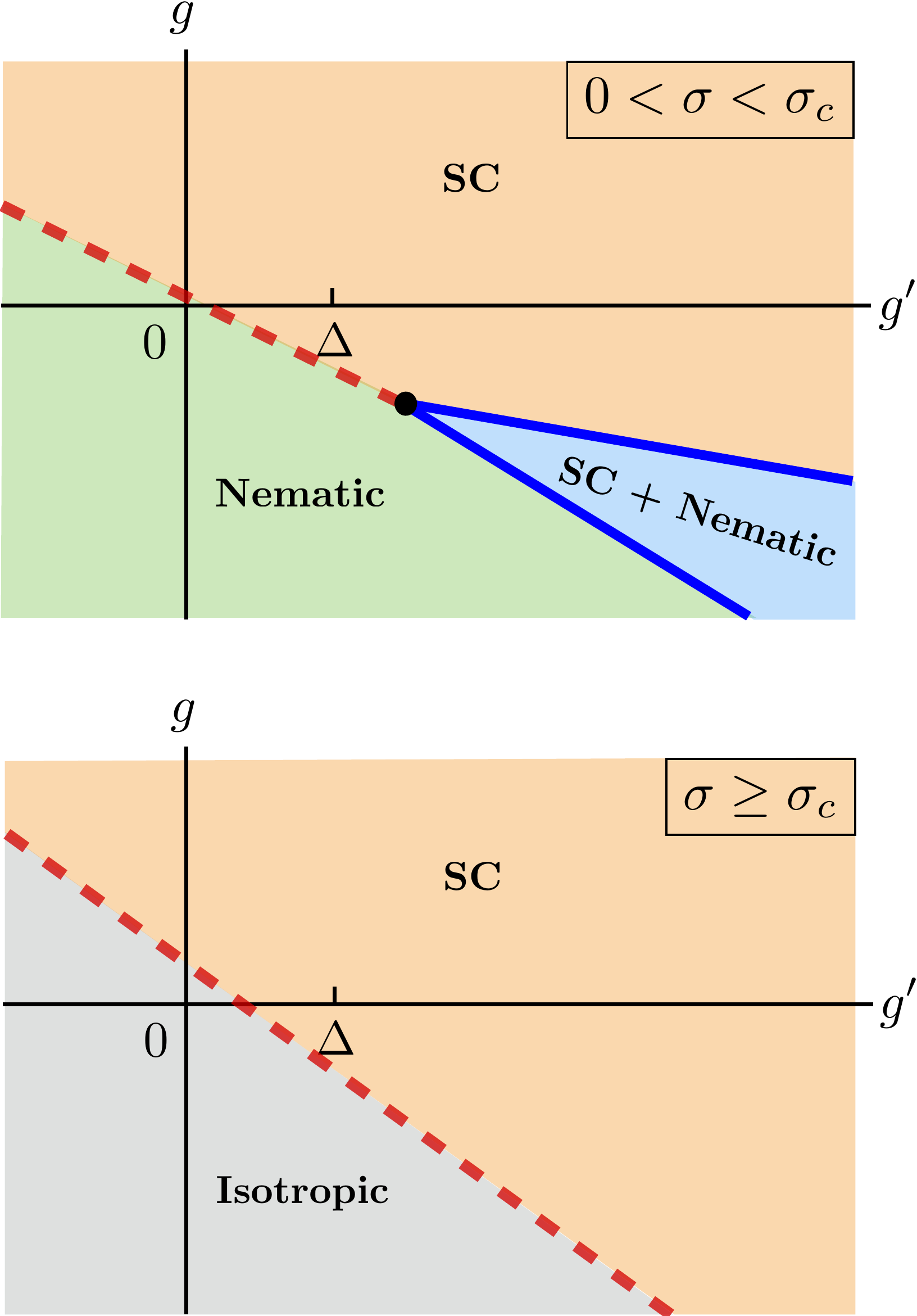}
	\caption{Phase diagram for the Hamiltonian in Eq.~\eqref{completeHamiltonian} at $T=0$ and  $\sigma>0$ (see text for definition of $\sigma_c$), 
with  $J' = J_z=0.01J$. Dashed (sketched by hand) and solid lines mark first- and second-order transitions respectively.}
	\label{fig:phasediagramcombined}
	\end{center}
\end{figure}
As mentioned in Fig.~\ref{fig:phasediagram+structurefactor}, for zero disorder there are three phases in the $g'-g$ phase diagram and a bi-critical point at $(g',g)=(\Delta,0)$. 
As shown in Fig~\ref{fig:phasediagramcombined}, for small but non-vanishing disorder, the stripe  (SC+stripe) phase is replaced by a nematic (SC+nematic) phase,
and the position of the bi-critical point shifts continuously.
There is a critical disorder strength $\sigma_\text{c}$ above which there is no nematic phase, and a first-order transition separates the SC and isotropic phases. 
(See Appendix~\ref{app:phasediagramzeroT} for a discussion of how the phase boundaries are determined in those plots.) \\


\subsection{Finite temperature}
\begin{figure}[t]
	\begin{center}
	\includegraphics[width=2.5in]{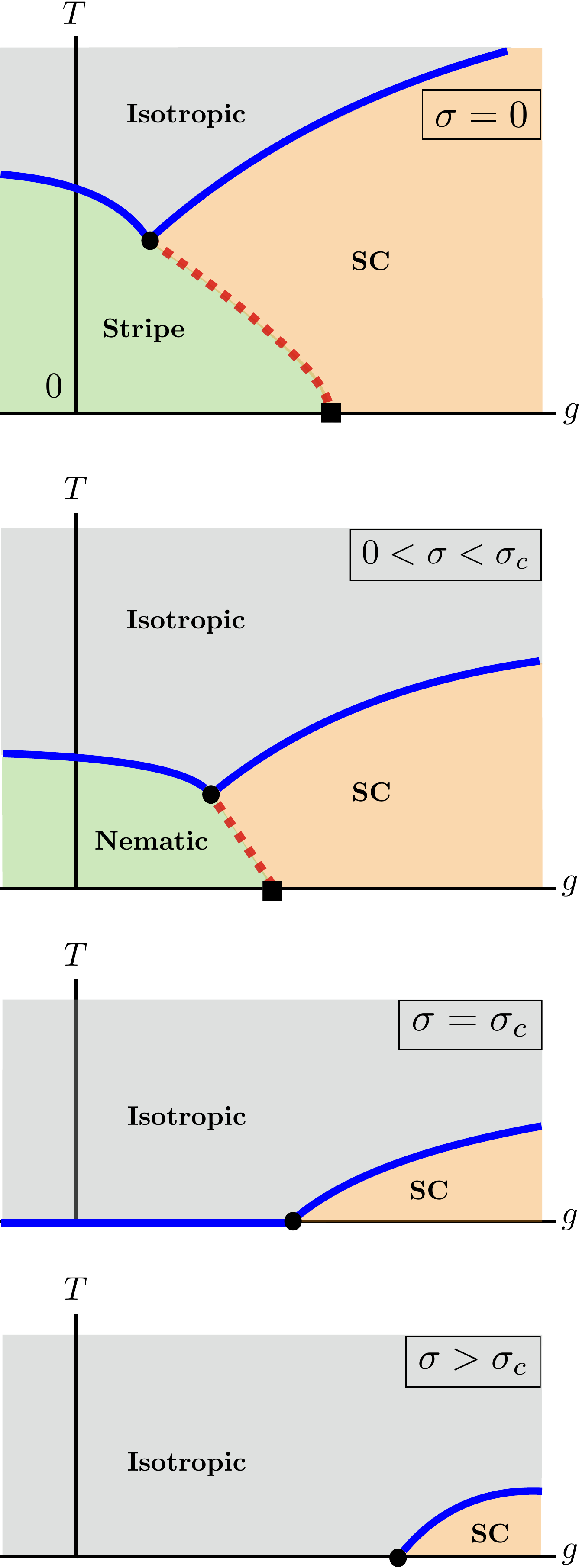}
	\caption{
	Evolution of the $T-g$ phase diagram for fixed $g' \ll \Delta$, as a function of $\sigma$ (see~\eqref{completeHamiltonian} for definitions of $g$ and $g'$).
Dashed (sketched by hand) 
and solid lines mark first- and second-order transitions, respectively. 
The bi-critical point moves toward large $g$ as disorder increases, whereas the zero-$T$ nematic-SC transition point behaves non-monotonically. }
	\label{fig:phasediagramfiniteleft}
	\end{center}
\end{figure}	

\begin{figure}[t]
	\begin{center}
	\includegraphics[width=2.5in]{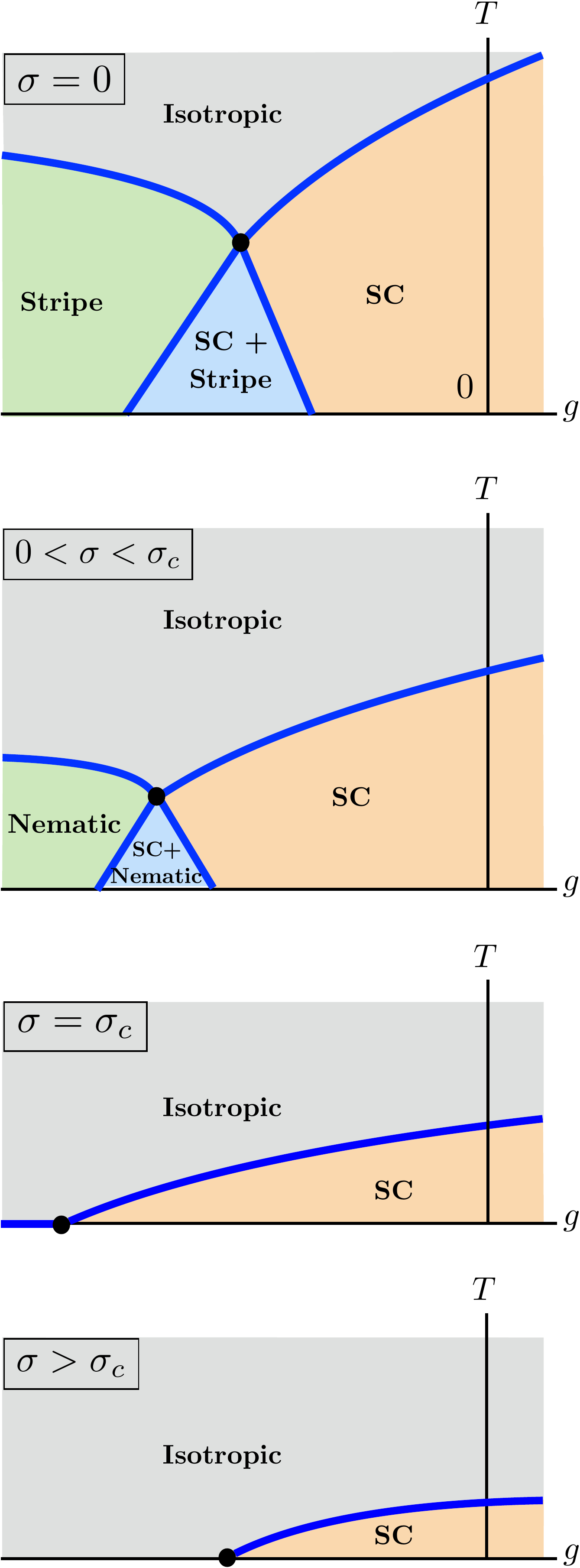}
	\caption{
	Evolution of the $T-g$ phase diagram for fixed $g' \gg \Delta$, as a function of $\sigma$  (see~\eqref{completeHamiltonian} for definitions of $g$ and $g'$).
	Solid lines mark second-order phase transitions. 
	The location of the tetra-critical point behaves non-monotonically as $\sigma$ increases, first moving towards smaller $g$ until $\sigma=\sigma_\text{c}$, and then moving towards larger~$g$. } 
	\label{fig:phasediagramfiniteright}
	\end{center}
\end{figure}
 
Consider the regimes of Fig.~\ref{fig:phasediagram+structurefactor} with $g' < \Delta$ and $g' > \Delta$, where $\Delta$ controls the tendency to break $C_4$ symmetry. We further limit our discussion to $g' \ll \Delta$ and $g' \gg \Delta$ to avoid the complication of a drifting bi-critical point in the presence of increasing disorder as indicated in Fig.~\ref{fig:phasediagramcombined}.
We plot in Figs.~\ref{fig:phasediagramfiniteleft} and~\ref{fig:phasediagramfiniteright} the $T-g$ phase diagram for fixed $g'$ and increasing disorder in these two regimes. Appendix~\ref{app:phasediagramfiniteT} gives detailed information about how phase boundaries and multicritical points are determined in these plots.

For $g' \ll \Delta$, the first-order transition between nematic and SC phases persists up to $\sigma_\text{c}$,
at which point the bi-critical point and the nematic phase disappear simultaneously.
As disorder is further increased beyond $\sigma_\text{c}$, the SC phase continues to exist, but this phase gets pushed steadily towards larger $g$. 
Quenched disorder tends to pin CDW locally, which indirectly suppresses the SC order. 
However, this effect is mitigated for larger $g$, at which point CDW order is suppressed and SC is favored. 

For $g' \gg \Delta$, the phase diagram has a tetra-critical point. 
At the critical disorder strength $\sigma_\text{c}$, this tetra-critical point and the nematic phase simultaneously vanish, similar to the situation of $g' \ll \Delta$.
Note that Fig.~\ref{fig:phasediagramfiniteleft} and Fig.~\ref{fig:phasediagramfiniteright} have the same $T_\text{multicritical}$ (temperatures corresponding to the multicritical points) and $\sigma_\text{c}$ (see Appendix~\ref{app:phasediagramfiniteT} for details). 

\section{Nematic correlation length} \label{sec:nematic}
\begin{figure}[h]
	\begin{center}
	\includegraphics{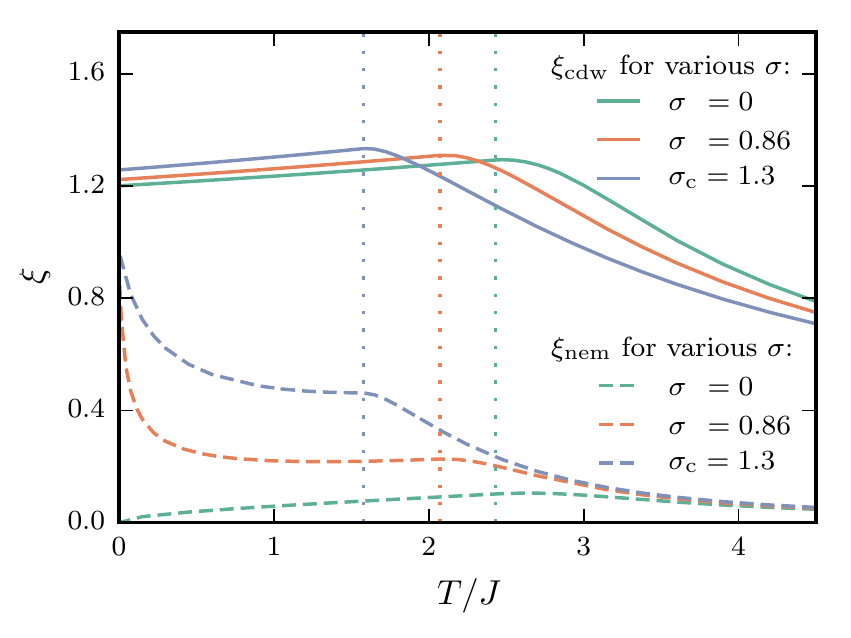}
	\caption{CDW and nematic correlation lengths with interlayer coupling $J_z=0.01$ and parameters given in Eq.~\eqref{input}. 
        These calculations use the large-$N$ saddle-point method.
	Dotted lines mark the SC transition temperatures. } 
	\label{fig:nemandCDW}
	\end{center}
\end{figure}

As shown in Sec.~\ref{subsec:region3}, when nematic order exists at $T=0$, the CDW structure factor $S_{\Phi_x}(\mathbf{k}=0)$ only reaches a maximum value at $T=0$. 
In other words, when the nematic correlation length $\xi_{\text{nem}} \to \infty$ we never observe $T_\text{max} > 0$. 
Here we turn to a regime where $T_\text{max}>0$ and no nematic phase occurs, and study the $\xi_\text{nem}$. Specifically, we choose input parameters corresponding to Region 1 in Fig.~\ref{fig:phasediagram+structurefactor}. 
As illustrated in Fig.~\ref{fig:nemandCDW}, $\xi_\text{nem} \ll \xi_\text{cdw}$ for weak disorder, where $\xi_{\text{cdw}}$ is the CDW correlation length. 
As the disorder strength increases, $\xi_\text{nem}$ grows and eventually becomes comparable with $\xi_\text{cdw}$ at low temperature. 
However, $\xi_\text{nem}$ still remains smaller than $\xi_\text{cdw}$. 
Implications of this result are discussed in Sec.~\ref{subsec:disscussion_exp} {\color{red}2}, and details of the calculation can be found in Appendix~\ref{app:nematiccorrelation}.


\section{Discussion} \label{sec:discussion}

\subsection{ Relation to experiments}
\label{subsec:disscussion_exp}

\noindent \textbf{1. X-ray scattering.} The idea of calculating CDW structure factors from a non-linear sigma model and comparing with X-ray data was initiated in Ref.~\onlinecite{hayward14}, where a model similar to Eq.~\eqref{completeHamiltonian} in a 2D, disorder-free system was shown to give good quantitative agreement with X-ray data.
However, two discrepancies remained. 
First of all, as $T \to 0$, the structure factor calculated using this model vanished, unlike what is seen in X-ray scattering experiments\cite{ghiringhelli12, chang12, achkar12, blackburn13, neto14, achkar14, blancocanosa14, hucker14}. 
Secondly, $T_\text{sc}$ was found to occur relatively far below the location of the CDW structure factor's maximum, whereas X-ray experiments find that $T_\text{sc} \ge T_{\text{max}}$\cite{ghiringhelli12, chang12, achkar12, blackburn13, achkar14, hucker14}. 
In our model, in the presence of quenched disorder ($\sigma \neq 0$), the CDW structure factor sustains a finite value at $T=0$ (due to  a pinning effect).
In addition, both the disorder and interlayer coupling present in our model prove to be effective for bringing $T_\text{sc}$ closer to $T_{\text{max}}$. \\

\noindent \textbf{2. Nematicity.} Macroscopically, long nematic correlations are observed experimentally in the pseudogap regime of several different cuprate materials~\cite{ando02,kivelson03, daou10, cyrchoiniere15, hinkov08, lawler10, fujita14,wu15}. 
These correlations could result from a nematic phase with infinite correlation length, or from a $C_4$-symmetric phase with strong nematic fluctuations and a finite correlation length that is long compared to $\xi_\text{cdw}$. 
The parameter regime that leads to the best agreement with X-ray data (Region 1 in Fig.~\ref{fig:phasediagram+structurefactor}), however, does not host a nematic phase, nor does it have a considerably longer nematic correlation length than $\xi_\text{cdw}$, as shown in Fig.~\ref{fig:nemandCDW}.
Meanwhile, in the regime we have studied where there is a low-$T$ nematic phase (Region 3 in Fig.~\ref{fig:phasediagram+structurefactor}), the CDW structure factor constantly increases as $T$ decreases, with no sign of turning down where the SC transition occurs. 
There are at least three possible explanations for these discrepancies:

\noindent \textbf{(i)} In our model we have taken $J_z=V_z$ for simplicity, but in reality the SC order below $T_{\text{sc}}$ is three-dimensional, whereas the CDW order always remains essentially two-dimensional (in low magnetic fields). 
It is possible that for $V_z \gg J_z$, where the 3D coupling makes the onset of SC order more robust and mean-field like, that a sharp depression of the CDW order could occur even where the nematic transition temperature is greater than the superconducting $T_{\text{sc}}$. 

\begin{figure}[t]
	\begin{center}
	\includegraphics{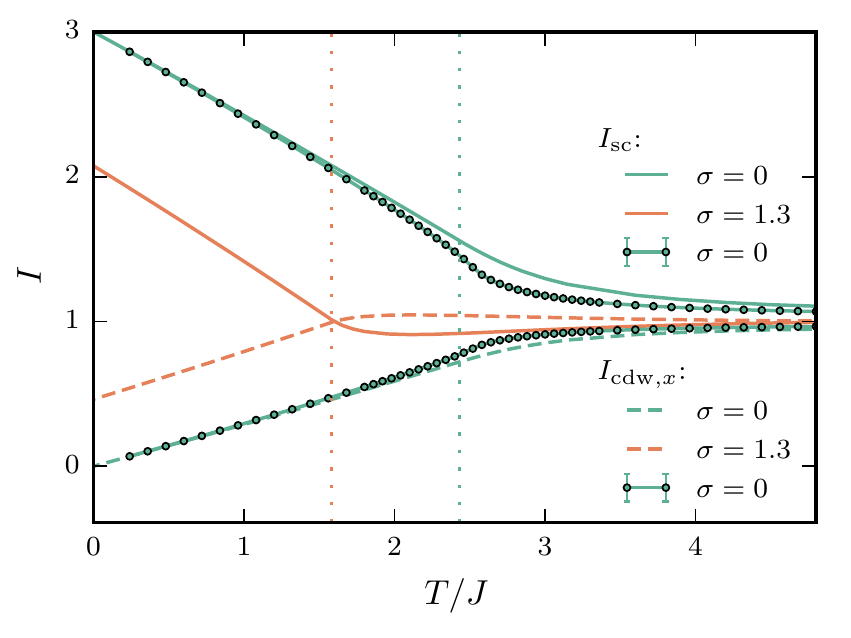}
	\caption{Integrated intensity from both large-$N$ mean-field (undecorated lines) and Monte Carlo (points with error bars) calculations. Dotted lines correspond to SC transitions. } 
	\label{fig:integrated}
	\end{center}
\end{figure}

\noindent \textbf{(ii)} Another limitation of our model comes from the constraint imposed in Eq.~\eqref{constraint}. This constraint is justified at temperatures well below any mean-field ordering temperature \cite{hayward14}.
However, at higher temperatures, the mean squared magnitudes of both the SC and the CDW order surely diminish. 
Specifically, these local magnitudes refer to the $\mathbf{k}$-integrated correlation functions,
\begin{align}
I_{\text{sc}} &\equiv \frac{1}{N} \int \frac{d^3\mathbf{k}}{(2 \pi)^3} \overline{ \langle   \Psi^{\dagger}(\mathbf{k}) \Psi(\mathbf{k}) \rangle}, \\
I_{\text{cdw,}x} &\equiv \frac{1}{N} \int \frac{d^3\mathbf{k}}{(2 \pi)^3} \overline{ \langle   \Phi^{\dagger}_x(\mathbf{k}) \Phi_x(\mathbf{k}) \rangle}.
\end{align}
and Eq.~\eqref{constraint} implies that $I_{\text{sc}} +I_{\text{cdw,}x}+ I_{\text{cdw,}y} = 3$, independent of $T$.
These integrated intensities are plotted in Fig.~\ref{fig:integrated} as functions of temperature for input parameters from Region 1 of Fig.~\ref{fig:phasediagram+structurefactor}. 
Here we see that,
{\it local} superconducting and CDW orders happily coexist at temperatures above their ordering temperatures. 
The competition between the two orders occurs predominantly as long-range correlations arise, at which point the system is forced to select one form of order or the other. 
In this sense, at least in our model, the competition does not primarily concern the amplitudes of the orders, which one might want to associate intuitively with a ``pairing scale'' in the case of superconductivity, but rather\cite{KFE} involves competition at the level of the helicity moduli, {\it i.e.} the ``superfluid stiffness'' in the case of superconductivity.

\noindent \textbf{(iii)} There is an implicit assumption in this discussion that the nematicity detected in experiment can be attributed to vestigial CDW order\cite{nie14}. 
However,  nematicity may have other origins in the pseudogap regime of cuprates, which is beyond the scope of our model.   This likely applies\cite{cyrchoiniere15} to the nematicity observed at somewhat lower  doping concentration in YBCO, {\it i.e.} for $\delta < 0.09$.

\subsection{Conclusions}
One aspect of our study with far-reaching implications is the remarkable degree to which the large-$N$ mean-field results qualitatively -- and in some cases even semi-quantitatively -- reproduce the Monte Carlo data at $N=2$.  
This observation was already apparent in Ref.~\onlinecite{hayward14}, but has now been extended to a wider range of circumstances.  
In particular, we have shown that this approach applies even in the presence of quenched disorder, where results obtained using this self-consistent mean-field theory combined with the replica trick reproduce the general trends seen in the disorder-averaged data from Monte Carlo simulations.  
Moving forward, this enables us to confidently use these approximate analytic approaches in other contexts.  For example, we can now more fully explore other effective field theories that could give rise to experimental features of the cuprates -- and other highly correlated materials with complex behavior -- related to fluctuating and intertwined orders.

\acknowledgements
We thank Adrian Del Maestro, David Hawthorn, Erica Carlson and Ryuichi Shindou for helpful discussions. 
This research was supported in part by Department of Energy under grant number DE-AC02-76SF00515 (L.N. and S.K.), the National Science Foundation under grant numbers DMR-1360789 (S.S.) and NSF PHY11-25915 (L.H., R.M. and S.S.). 
Support was also provided by the Natural Sciences and Engineering Research Council of Canada, the Perimeter Institute for Theoretical Physics, the John Templeton Foundation, and the Canada Research Chair program.
Research at Perimeter Institute is supported by the Government of Canada through Industry Canada 
and by the Province of Ontario through the Ministry of Research and Innovation. 
L.N. thanks the support from the Stanford Graduate Fellowship.
Monte Carlo simulations were performed using SHARCNET high performance computing resources.

\begin{widetext}
\appendix
\onecolumngrid

\section{Parameter conversion}
\label{app:parameterConversion}
\renewcommand{\theequation}{A.\arabic{equation}}
We provide here the conversion rules between our model (Eq.~\eqref{completeHamiltonian}) and the model used in Refs.~\onlinecite{hayward14}~and~\onlinecite{diamag14}. The Hamiltonian used in these previous studies is
\begin{align}
\widetilde{H} &= -  \sum\limits_{\langle ij \rangle_{xy}} \left[ 
\sum\limits_{\alpha=1}^2 n_{i\alpha}n_{j\alpha}  
+ \lambda \sum\limits_{\alpha=3}^6 n_{i\alpha}n_{j\alpha} \right] 
- \sum\limits_{\langle ij \rangle_z}  \left[   \widetilde{V}_z\sum\limits_{\alpha=1}^2 n_{i\alpha}n_{j\alpha}  
+ \widetilde{J}_z   \sum\limits_{\alpha=3}^6 n_{i\alpha}n_{j\alpha}   \right]   \nonumber\\
& \quad {}+ \frac{\widetilde{g} + 4 (\lambda -1)}{2} \sum\limits_i \sum\limits_{\alpha=3}^6 n^2_{i\alpha} 
+ \frac{\widetilde{g}'}{2} \sum\limits_i \left( \sum\limits_{\alpha=3}^6 n^2_{i\alpha}   \right)^2
+  \frac{w}{2}  \sum\limits_i   \left[  (n^2_{i3} + n^2_{i4})^2 +   (n^2_{i5} + n^2_{i6})^2   \right]
  + \frac{1}{2}\sum\limits_i\sum\limits_{\alpha=3}^6 \widetilde{h}_{i\alpha} n_{i\alpha}, 
\label{HS}
\end{align}
where $n_1 + i n_2$ represents the SC order parameter, and $n_3 + i n_4$ and $n_5 + i n_6$ represent the two CDW order parameters. 
This model imposes the constraint $\sum_{\alpha=1}^6 n_{i\alpha}^2 = 1$.
Note that Refs.~\onlinecite{hayward14} and~\onlinecite{diamag14} do not include the effects of interlayer coupling and disorder, so that $\widetilde{V}_z = \widetilde{J}_z = \widetilde{h} = 0$ in these references. 
One can write the fields $\Psi$, $\Phi_{x}$ and $\Phi_{y}$ from Eq.~\eqref{completeHamiltonian} in discretized form in terms of the components $n_{i \alpha}$ as
\begin{align}
\Psi_i &= \sqrt{3N} \times \left( n_{i1}, n_{i2}, \ldots ,n_{iN} \right)^T,  \nonumber\\
\Phi_{xi} &= \sqrt{3N} \times \left( n_{i,N+1}, n_{i,N+2}, \ldots ,n_{i,2N} \right)^T,   \nonumber\\
\Phi_{yi} &= \sqrt{3N} \times \left( n_{i,2N+1}, n_{i,2N+2}, \ldots ,n_{i,3N} \right)^T,  \nonumber\\
h_{i}  &=   \frac{\sqrt{3N}}{2} \times \left( \widetilde{h}_{i,N+1}, \widetilde{h}_{i,N+2}, \ldots , \widetilde{h}_{i,3N} \right)^T,
\label{redefine}
\end{align}
where $N$ is the number of components of each order parameter; 
$N=2$ in Refs.~\onlinecite{hayward14} and~\onlinecite{diamag14} as well as in the Monte Carlo calculations within this paper.
With the transformation defined in Eq.~\eqref{redefine}, it is then straightforward to show that the conversion rules between the parameters in Eqs.~\eqref{completeHamiltonian} and~\eqref{HS} are
\be
\widetilde{T} = T/6, \:\: \lambda = J/K, \:\:  \widetilde{J}_z = J_z,  \:\:  \widetilde{V}_z = V_z, \:\: \widetilde{g} =  g -4 (J/K-1), \:\: \widetilde{g}'  = 3(g'+\Delta), \:\:
w=-6\Delta,
\label{conversion}
\ee
where $T$ and $\widetilde{T}$ are the temperatures corresponding to Eqs.~\eqref{completeHamiltonian} and~\eqref{HS}, respectively. 

\section{Effect of $1/N$ correction}\label{app:oneoverN}
$1/N$ corrections do not change high and low temperature behaviors of the CDW structure factor $S_{\Phi_x}(\mathbf{k}=0)$, but these corrections do have a quantitative effect on its maximum. In Fig.~\ref{fig:correctionComparison}, we compare Monte Carlo results for $S_{\Phi_x}(\mathbf{k}=0)$ to large-$N$ results with and without $1/N$ correction for two different sets of parameters.
\begin{figure}[h]
	\begin{center}
	\includegraphics{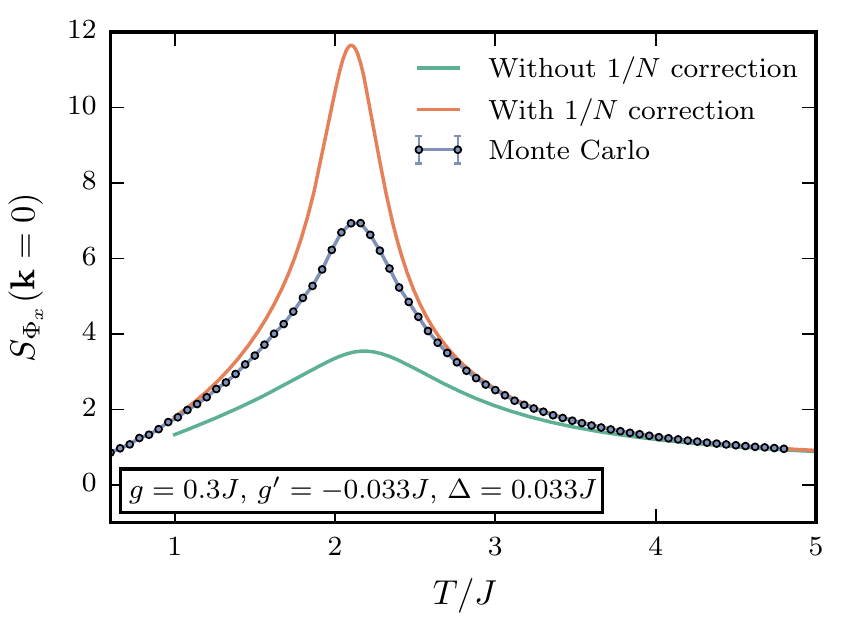}
	\hspace{1em}
	\includegraphics{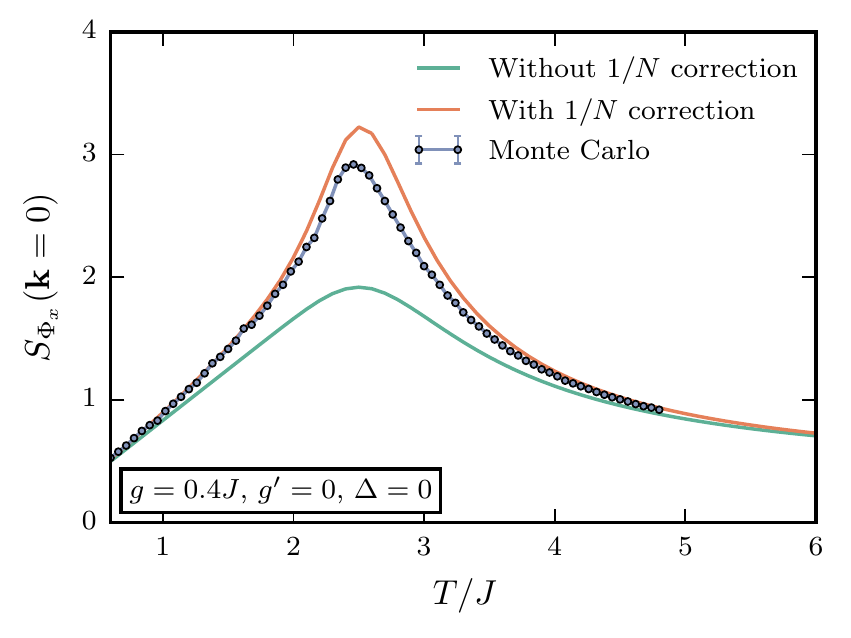}
	\caption{The effect of $1/N$ corrections in comparion with Monte Carlo data for two different sets of parameters. Both parameters sets have $K=J$, $V_z = J_z = 0$, $J'=0$ and $\sigma=0$. The parameters for both plots correspond to Region~1 of Fig.~\ref{fig:phasediagram+structurefactor}.} 
	\label{fig:correctionComparison}
	\end{center}
\end{figure}
In the first plot, we find a large effect of the $1/N$ corrections on the height of the peak: this is the region where the inverse propagator of 
$\Phi$ is the smallest, and so small ``self-energy'' corrections can have a large effect on the correlator. In the second plot, we see that the $1/N$ correction greatly improves the agreement with Monte Carlo.

\section{Details of Section~\ref{sec:phaseDiagram}}
\subsection{Zero temperature}
\label{app:phasediagramzeroT}
\renewcommand{\theequation}{C.\arabic{equation}}
For $T=0$ and $\sigma=0$, only the on-site terms in Eq.~\eqref{completeHamiltonian} remain and we are left with a spatially homogeneous ground state that satisfies
\be
\frac{H}{V} = g\bigg[|\Phi_x|^2 + |\Phi_y|^2 \bigg]    -\frac{\Delta}{N} \bigg[ |\Phi_x|^2 - |\Phi_y|^2 \bigg]^2  + \frac{g'}{N}  \bigg[  |\Phi_x|^2+ |\Phi_y|^2\bigg]^2, 
\ee
where $V$ is the volume of the system, and we have set $J=K$ so that the nearest-neighbour terms in Eq.~\eqref{completeHamiltonian} become constants. The problem amounts to searching for minima of a two-variable function 
\be
H(x,y) =  g (x+y) -\frac{\Delta}{N} (x-y)^2 + \frac{g'}{N} (x+y)^2,
\ee
with $\Delta>0$ and constraints
\be
x \ge 0, \quad y \ge 0, \quad x+y \le 3N.
\ee
The results are (as shown in Fig.~\ref{fig:phasediagram+structurefactor}):
\begin{itemize}[leftmargin=*]
\item
When $g'>\Delta$,
\be
H(x,y) _{\min}=
\left\{ \begin{array}{lll}   
H(0,0) &  \text{ (SC)} &  \text{for } g>0 \\
H(-gN/[2(g'-\Delta)],0) & \text{ (stripe+SC)} & \text{for } 6(\Delta-g') < g \le 0 \\
H(3N,0) &  \text{ (stripe)} & \text{for } g \le 6(\Delta-g').  \end{array} \right.
\ee
\item
When $g' \le \Delta$,
\be
H(x,y) _{\min}=
\left\{ \begin{array}{lll}   
H(3N,0) &  \text{ (stripe)} &  \text{for }  g<3(\Delta-g') \\
H(0,0) &  \text{ (SC)} &  \text{for } g \ge 3(\Delta-g'). \end{array} \right.
\ee
\end{itemize}

For $T=0,$ and $\sigma \ne 0$ (Fig.~\ref{fig:phasediagramcombined}) we must numerically solve Eqs.~\eqref{saddle1},~\eqref{saddle2}~and~\eqref{saddle3} for the $g'$-dependence of $g$, under the following conditions (where $-4J-2J_z-g-4g'+\eta+\mu$ is the mass term of SC order parameter):
\begin{itemize}[leftmargin=*]
\item
Nematic to SC$+$nematic: \\
$-4J-2J_z-g-4g'+\eta+\mu=0, \quad \tilde m^2=0, \quad T=0;\quad$ solve~\eqref{saddle1}, ~\eqref{saddle2} and~\eqref{saddle3}.
\item
SC to SC$+$nematic: \\
$-4J-2J_z-g-4g'+\eta+\mu=0, \quad \mathcal{N}=0, \quad T=0; \quad$ solve~\eqref{saddle1},~\eqref{saddle3}, and 
$\partial / \partial \mathcal{N}$ over both sides of~\eqref{saddle2} with $\mathcal{N}=0$. 
\end{itemize}

\subsection{Finite temperature}
\label{app:phasediagramfiniteT}
\renewcommand{\theequation}{C.\arabic{equation}}

The second-order phase transitions in Figs.~\ref{fig:phasediagramfiniteleft}~and~\ref{fig:phasediagramfiniteright} are obtained by numerically solving Eqs.~\eqref{saddle1},~\eqref{saddle2}~and~\eqref{saddle3} under corresponding conditions:

\begin{itemize}[leftmargin=*]
\item
Isotropic to SC: \\
$-4J-2J_z-g-4g'+\eta+\mu=0,\quad  \mathcal{N}=0, \quad \tilde m^2=0,\quad $ solve~\eqref{saddle1}~and~\eqref{saddle3}.
\item
Isotropic to stripe (isotropic to nematic if $\sigma \ne 0$): \\
$\mathcal{N}=0,\quad \tilde m^2=0,\quad$ solve~\eqref{saddle1},~\eqref{saddle3}~and~$\partial / \partial \mathcal{N}$ over both sides of ~\eqref{saddle2} with $\mathcal{N}=0.$
\item
Stripe to SC$+$stripe (nematic to SC$+$nematic if $\sigma \ne 0$): \\$-4J-2J_z-g-4g'+\eta+\mu=0, \quad \tilde m^2=0,\quad$ solve~\eqref{saddle1}, ~\eqref{saddle2} and~\eqref{saddle3}.
\item
SC to SC$+$stripe (SC$+$nematic if $\sigma \ne 0$): \\
$-4J-2J_z-g-4g'+\eta+\mu=0, \quad \mathcal{N}=0,\quad$ solve~\eqref{saddle1},~\eqref{saddle3}~and~$\partial / \partial \mathcal{N}$ over both sides of ~\eqref{saddle2} with $\mathcal{N}=0.$
\end{itemize}

To find the multicritical points in Figs.~\ref{fig:phasediagramfiniteleft} and~\ref{fig:phasediagramfiniteright}, we must impose the conditions
\be
-4J-2J_z-g-4g'+\eta+\mu=0,\quad  \mathcal{N}=0, \quad \tilde m^2=0,\quad 
\ee
and solve~\eqref{saddle1},~\eqref{saddle3}~and~$\partial / \partial \mathcal{N}$ over both sides of ~\eqref{saddle2} with $\mathcal{N}=0.$

The reason $T_{\text{multicritical}}$ is independent of $g$ and $g'$ is that Eq.~\eqref{saddle3} is decoupled from Eqs.~\eqref{saddle1} and~\eqref{saddle2} at the multicritical point. At $\sigma=\sigma_\text{c}$ the situation is similar: at $T=0$ Eq.~\eqref{saddle3} becomes trivial and we only need to solve Eqs.~\eqref{saddle1} and~\eqref{saddle2}.

\section{Nematic correlation function}\label{app:nematiccorrelation}
\renewcommand{\theequation}{D.\arabic{equation}}
We present some technical aspects of computing nematic correlation functions using the replica trick. 
Neglecting the terms related to superconductivity  ($\eta$ and $\Psi$), we can rewrite the replicated Hubbard-Stratonovich Hamiltonian of Eq.~\eqref{replicaHSHamiltonian} in $k$-space as
\begin{align}
H_{\text{replica}} &= \int_{d\mathbf{k}}  \sum\limits_{a} \Bigg\{ A(\mathbf{k})\Phi_{a,x}^{\dagger}(\mathbf{k}) \Phi_{a,x}(\mathbf{k}) + B(\mathbf{k}) \Phi_{a,y}^{\dagger}(\mathbf{k}) \Phi_{a,y}(\mathbf{k})   \nonumber\\
&\quad{}+  \mathcal{N}_a(\mathbf{k}) \int_{d\mathbf{p}}\Big[ \Phi_{a,x}^{\dagger} (\mathbf{k+p}) \Phi_{a,x}(\mathbf{p}) - \Phi_{a,y}^{\dagger} (\mathbf{k+p})\Phi_{a,y}(\mathbf{p})  \Big]  + \frac{\mathcal{N}_a(\mathbf{k}) \mathcal{N}_a(-\mathbf{k}) N}{4\Delta} \Bigg\} \nonumber\\
&+  \int_{d\mathbf{k}}  \sum\limits_{a,b}  \left(-\frac{2\sigma^2}{T}\right)\Phi^{\dagger}_{a}(\mathbf{k}) \Phi_{b}(\mathbf{k}),
\end{align}
where $\int_{d\mathbf{k}} \equiv \int \frac{d^3\mathbf{k}}{(2\pi)^3}$, and 
\be
A, B(\mathbf{k}) \equiv -2J(\cos k_x + \cos k_y) \mp 2J'(\cos k_x-\cos k_y) -2V_z\cos k_z+\mu.
\label{ABdefinition}
\ee
Our goal is to compute $\langle \mathcal{N}_a(\mathbf{k}) \mathcal{N}_a(\mathbf{-k})   \rangle$ in the limit $M\to 0$, where $M$ is the total number of replicas. First, we integrate out $\Phi$ and obtain an effective Hamiltonian for $\mathcal{N}$, which satisfies
\begin{align}
&e^{-\beta H_\text{eff}[\mathcal{N}]}  =\int \mathcal{D} \Phi \mathcal{D} \Phi^{\dagger} 
\exp\Big\{ -\beta H_{\text{replica}} \Big\} \nonumber\\
&= \int \mathcal{D} \Phi  \mathcal{D} \Phi^{\dagger} \exp \Bigg\{ 
\! -\beta \! \int_{d\mathbf{k}} \! \left(\Phi_{1,x}^{\dagger}, \ldots , \Phi_{M,x}^{\dagger} \right) A_{\sigma}(\mathbf{k}) \left(\Phi_{1,x}, \ldots ,\Phi_{M,x}\right)^T   +   \left(\Phi_{1,y}^{\dagger}, \ldots ,\Phi_{M,y}^{\dagger} \right) B_{\sigma}(\mathbf{k}) \left(\Phi_{1,y}, \ldots ,\Phi_{M,y}\right)^T\Bigg\}   \nonumber\\
& \qquad \qquad \qquad \times    \exp\Bigg\{  
\! -\beta \sum\limits_a \int_{d\mathbf{k} d\mathbf{p}} \mathcal{N}_a(\mathbf{k}) \Big[ \Phi_{a,x}^{\dagger} (\mathbf{k+p}) \Phi_{a,x}(\mathbf{p}) - \Phi_{a,y}^{\dagger} (\mathbf{k+p})\Phi_{a,y}(\mathbf{p})  \Big]  + \frac{\mathcal{N}_a(\mathbf{k}) \mathcal{N}_a(-\mathbf{k}) N}{4\Delta}  \Bigg\} , 
\label{integrateoutPhi}
\end{align}
where 
\be
A_{\sigma}, B_{\sigma}(\mathbf{k}) \equiv A,B(\mathbf{k}) \cdot \mathbb{I}_{M\times M} + \left(-\frac{2\sigma^2}{T} \right)  \left( \begin{array}{ccc}  
1& \ldots & 1  \\
\vdots & \ddots & \vdots \\
1 & \ldots & 1
\end{array} \right).
\ee
We now expand the last exponential term in Eq.~\eqref{integrateoutPhi} to quadratic order in $\mathcal{N}$ (which is equivalent to keeping the diagrams up to order $1/N$, as shown in Fig.~\ref{fig:diagrams}), and arrive at
\begin{align}
& \exp\Bigg\{  -\beta \sum\limits_a \int_{d\mathbf{k} d\mathbf{p}} \mathcal{N}_a(\mathbf{k}) \Big[ \Phi_{a,x}^{\dagger} (\mathbf{k+p}) \Phi_{a,x}(\mathbf{p}) - \Phi_{a,y}^{\dagger} (\mathbf{k+p})\Phi_{a,y}(\mathbf{p})  \Big]  \Bigg\} \nonumber\\
&\qquad = 1 - \beta \sum\limits_a \int_{d\mathbf{k} d\mathbf{p}} \mathcal{N}_a(\mathbf{k}) \Big[ \Phi_{a,x}^{\dagger} (\mathbf{k+p}) \Phi_{a,x}(\mathbf{p}) - \Phi_{a,y}^{\dagger} (\mathbf{k+p})\Phi_{a,y}(\mathbf{p})  \Big]  \nonumber\\
&\quad \qquad{} + \frac{\beta^2}{2}   \sum\limits_{a,b}   \int_{d\mathbf{k} d\mathbf{p}  d\mathbf{k'} d\mathbf{p'}} \mathcal{N}_a({\mathbf{k}}) \mathcal{N}_b({\mathbf{k'}})  \nonumber\\
& \qquad \qquad \qquad   \cdot   \Big[ \Phi_{a,x}^{\dagger}(\mathbf{k+p}) \Phi_{a,x}(\mathbf{p}) -  \Phi_{a,y}^{\dagger}(\mathbf{k+p}) \Phi_{a,y}(\mathbf{p})   \Big]   \cdot \Big[ \Phi_{b,x}^{\dagger}(\mathbf{k'+p'}) \Phi_{b,x}(\mathbf{p'}) -  \Phi_{b,y}^{\dagger}(\mathbf{k'+p'}) \Phi_{b,y}(\mathbf{p'})   \Big]     
\label{expansion}
\end{align}

\begin{figure}[t]
	\begin{minipage}{1\textwidth}
	\begin{center}
	\includegraphics[width=6cm]{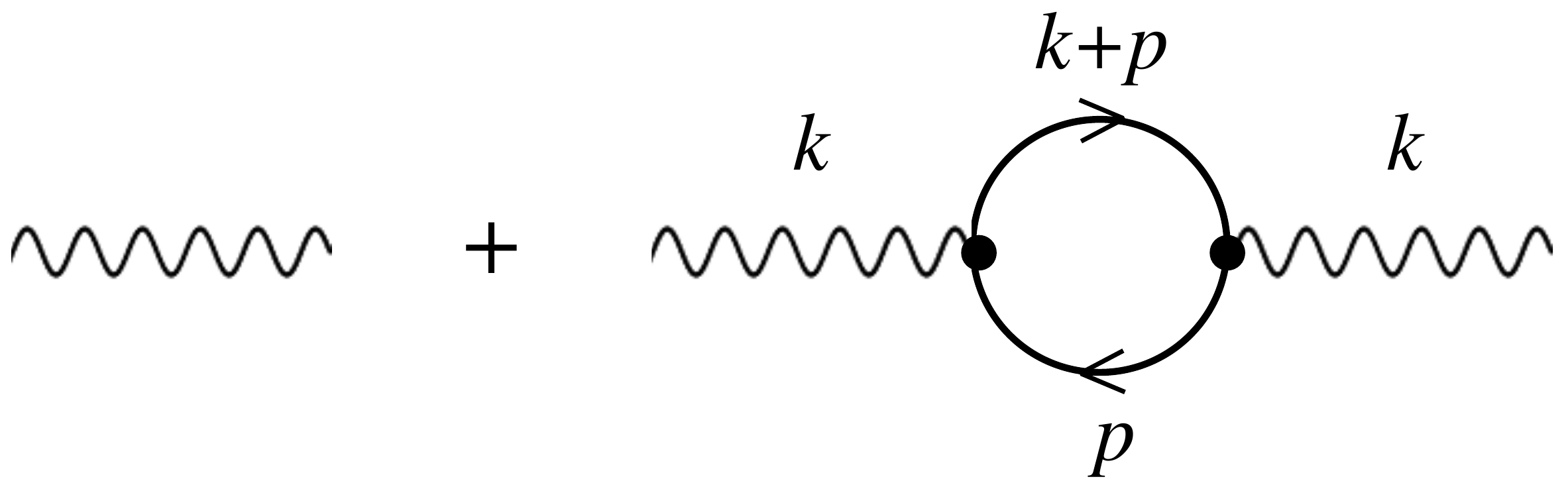}
	\caption{The lowest order diagrams corresponding to Eq.~\eqref{integrateoutPhi} (both are order $1/N$). The wavy lines correspond to the bare propagator of $\mathcal{N}$, and the solid lines correspond to the bare propagator of $\Phi_{x(y)}$. Each vertex is order 1 and the loop is of order $N$ (coming from the sum of all the $\Phi$'s). }
	\label{fig:diagrams}
	\end{center}
	\end{minipage}
\end{figure}

\noindent The term linear in $\beta$ will vanish due to the cancellation between $\Phi_x$ and $\Phi_y$, and we will drop the constant 1. The  remaining $\beta^2$ term gives 
\be
\beta H_\text{eff}[\mathcal{N}] = \sum\limits_{a,b}  \int_{d\mathbf{k}} \mathcal{N}_a(\mathbf{k}) t_{ab}(\mathbf{k}) \mathcal{N}_b(\mathbf{-k}),
\label{Heff}
\ee
\be
t_{ab}(\mathbf{k}) \equiv\frac{\delta_{ab}}{4\Delta T} - \frac{1}{2T^2}\int_{d\mathbf{p}} \langle \Phi_{a,x}^{\dagger} \Phi_{b,x}(\mathbf{k+p}) \rangle  \langle  \Phi_{b,x}^{\dagger} \Phi_{a,x}(\mathbf{p}) \rangle   + \langle \Phi_{a,y}^{\dagger} \Phi_{b,y}(\mathbf{k+p})  \rangle  \langle  \Phi_{b,y}^{\dagger} \Phi_{a,y}(\mathbf{p}) \rangle ,
\label{tab}
\ee
where we have included $\frac{\mathcal{N}_a(\mathbf{k}) \mathcal{N}_a(\mathbf{-k}) N}{4\Delta}$ from Eq.~\eqref{integrateoutPhi}, and dropped the factor of $N$.
The nematic correlation function is
\be
\langle \mathcal{N}_a(\mathbf{k} ) \mathcal{N}_a(\mathbf{-k})  \rangle =  \frac{1}{M} \sum\limits_b \langle \mathcal{N}_b(\mathbf{k} ) \mathcal{N}_b(\mathbf{-k})  \rangle   = \frac{1}{M} (s^{-1}_1+ \ldots + s^{-1}_M) 
\label{NNcorrelation}
\ee
where $\{ s_a \}$ are  eigenvalues of $2t_{ab}$ (the factor 2 is due to the fact that the $\Phi's$ are complex). Our strategy is to first compute $\langle \Phi^{\dagger}_{a,x} \Phi_{b,x}(\mathbf{k})\rangle$ in Eq.~\eqref{tab} by diagonalizing $A_{\sigma}(\mathbf{k})$, which gives
\be
\langle  \Phi_{a,x}^{\dagger} \Phi_{b,x}(\mathbf{k}) \rangle = (\delta_{ab} - \frac{1}{M}) \frac{T}{A(\mathbf{k})} + \frac{1}{M} \cdot \frac{T}{A(\mathbf{k}) - 2\sigma^2 M/T}.
\ee
We then diagonalize $t_{ab}$ and obtain $\langle \mathcal{N}_a(\mathbf{k} ) \mathcal{N}_a(\mathbf{-k})  \rangle$ according to Eq.~\eqref{NNcorrelation}, which yields
\begin{align}
&\langle \mathcal{N}_a(\mathbf{k})  \mathcal{N}_a(\mathbf{-k})  \rangle (M\to 0)  \nonumber\\
&\; =\frac{ \frac{1}{2\Delta T} - \int_{d\mathbf{p}} \Big\{   \frac{1}{A(\mathbf{k+p})A(\mathbf{p})}   + \frac{2\sigma^2}{T}  \left[ \frac{1}{A^2(\mathbf{k+p})A(\mathbf{p})} + \frac{1}{A(\mathbf{k+p})A^2(\mathbf{p})} \right]   -  \frac{4\sigma^4}{T^2} \int_{d\mathbf{p}}    \frac{1}{A^2(\mathbf{k+p})A^2(\mathbf{p})}  + A \leftrightarrow B \Big\}     }{    \Bigg\{  \frac{1}{2\Delta T} - \int_{d\mathbf{p}} \Big\{   \frac{1}{A(\mathbf{k+p})A(\mathbf{p})}   + \frac{2\sigma^2}{T}  \left[ \frac{1}{A^2(\mathbf{k+p})A(\mathbf{p})} + \frac{1}{A(\mathbf{k+p})A^2(\mathbf{p})} \right]   + A \leftrightarrow B \Big\}     \Bigg\}^2 } . \nonumber\\
&&
\label{nonzerokresult} 
\end{align}
At $\mathbf{k}=0$, Eq.~\eqref{nonzerokresult} reduces to
\be
\langle \mathcal{N}_a(0)  \mathcal{N}_a(0)   \rangle (M\to 0)  = 
\frac{ \frac{1}{2\Delta T}  - \int_{d\mathbf{p}} \left(  \frac{1}{A^2(\mathbf{p})} + \frac{1}{B^2(\mathbf{p})}  +\frac{4\sigma^2}{TA^3(\mathbf{p})} +\frac{4\sigma^2}{TB^3(\mathbf{p})}   \right)  + \int_{d\mathbf{p}} \left( \frac{4\sigma^4}{T^2A^4(\mathbf{p})} + \frac{4\sigma^4}{T^2B^4(\mathbf{p})} \right)}{ \left[ \frac{1}{2\Delta T}  - \int_{d\mathbf{p}} \left(  \frac{1}{A^2(\mathbf{p})} + \frac{1}{B^2(\mathbf{p})}  +\frac{4\sigma^2}{TA^3(\mathbf{p})} +\frac{4\sigma^2}{TB^3(\mathbf{p})}   \right)\right]^2 },
 \label{zerokresult}
\ee
where $A$ and $B$ are defined in Eq.~\eqref{ABdefinition}. A nematic transition occurs when the denominator vanishes such that
\be
 \frac{1}{2\Delta T}  = \int_{d\mathbf{p}} \left(  \frac{1}{A^2(\mathbf{p})} + \frac{1}{B^2(\mathbf{p})}  +\frac{4\sigma^2}{TA^3(\mathbf{p})} +\frac{4\sigma^2}{TB^3(\mathbf{p})}   \right),
\ee
which is consistent with mean-field saddle-point equation~\eqref{saddle2} after taking a derivative with respect to $\mathcal{N}$ and setting $\mathcal{N}=0$ in~\eqref{saddle2}.

\end{widetext}


\bibliographystyle{h-physrev}
\bibliography{References}

\end{document}